\documentclass[conference]{IEEEtran}
\IEEEoverridecommandlockouts

\usepackage{cite}
\usepackage{amsmath,amssymb,amsfonts}
\usepackage{textcomp}
\usepackage{xcolor}
\def\BibTeX{{\rm B\kern-.05em{\sc i\kern-.025em b}\kern-.08em
    T\kern-.1667em\lower.7ex\hbox{E}\kern-.125emX}}

\usepackage{balance}
\usepackage{mathrsfs}
\usepackage{mathtools}
\usepackage[english]{babel}
\usepackage{float}
\usepackage[pdftex]{graphicx}
\usepackage{epstopdf}
\usepackage{color}
\usepackage{hyperref}
\usepackage{multirow}
\usepackage{graphicx}
\usepackage{bm}
\usepackage{balance}
\usepackage{microtype}
\usepackage[caption=false,font=footnotesize]{subfig}
\usepackage{amsthm, thmtools}
\usepackage{mathtools}
\usepackage[most]{tcolorbox}

\usepackage{tikz,lipsum}
\usepackage[most]{tcolorbox}

\newtcolorbox{Box1}[2][]{
                lower separated=false,
                colback=white,
colframe=white!20!gray,fonttitle=\bfseries,
colbacktitle=white!10!gray,enhanced,
attach boxed title to top left={xshift=0.1cm,
        yshift=-0.01mm}, 
title=#2}
\newtheorem{theorem}{Theorem}
\makeatletter
\newcommand{\settheoremtag}[1]{
  \let\oldthetheorem\thetheorem
  \renewcommand{\thetheorem}{#1}
  \g@addto@macro\endtheorem{
    \addtocounter{theorem}{-1}
    \global\let\thetheorem\oldthetheorem}
  }
\makeatother

\usepackage{algorithm}  
\usepackage{algpseudocode}  
\usepackage{multirow}
\usepackage{rotating} 
\usepackage{makecell}
\usepackage{array}
\usepackage{tabularray}
\usepackage{pifont}
\usepackage{amssymb}
\usepackage{pifont}
\newcommand{\cmark}{\ding{51}} 
\newcommand{\xmark}{\ding{55}} 

\newcommand{\profe}[1]{{\color{black}{#1}}}

\usepackage{bm}

\begin{document}

\title{Deep Distillation Gradient Preconditioning \\ for Inverse Problems\\
}
\author{
\IEEEauthorblockN{Romario Gualdr\'on-Hurtado$^{\star}$, Roman Jacome$^{\dagger}$, Leon Suarez$^{\star}$, Laura Galvis$^{\star}$, Henry Arguello$^{\star}$}
\IEEEauthorblockA{$^{\star}$Department of Systems and Informatics Engineering
\\
$^{\dagger}$Department of Electrical, Electronics, and Telecommunications Engineering
\\
Universidad Industrial de Santander, Bucaramanga, Colombia, 680002}
}

\maketitle

\begin{abstract}
Imaging inverse problems are commonly addressed by minimizing measurement consistency and signal prior terms. While huge attention has been paid to developing high-performance priors, even the most advanced signal prior may lose its effectiveness when paired with an ill-conditioned sensing matrix that hinders convergence and degrades reconstruction quality. In optimization theory, preconditioners allow improving the algorithm's convergence by transforming the gradient update. Traditional linear preconditioning techniques enhance convergence, but their performance remains limited due to their dependence on the structure of the sensing matrix. Learning-based linear preconditioners have been proposed, but they are optimized only for data-fidelity optimization, which may lead to solutions in the null-space of the sensing matrix. This paper employs knowledge distillation to design a nonlinear preconditioning operator. In our method, a teacher algorithm using a better-conditioned (synthetic) sensing matrix guides the student algorithm with an ill-conditioned sensing matrix through gradient matching via a preconditioning neural network. We validate our nonlinear preconditioner for plug-and-play FISTA in single-pixel, magnetic resonance, and super-resolution imaging tasks, showing consistent performance improvements and better empirical convergence.
\end{abstract}

\vspace{-0.5em}

\section{Introduction} \label{sec:intro}
Inverse problems are at the core of many imaging applications, such as super-resolution, magnetic resonance imaging, and single-pixel cameras \cite{yang2010image,pruessmann1998coil,duarte2008single,gualdron2025improving}, where an unknown signal $\mathbf{x}\in\mathbb{R}^n$ is recovered from incomplete and noisy measurements $\mathbf{y}\in\mathbb{R}^m$ (with $m\ll n$) modeled as $ \mathbf{y} = \mathbf{H}\mathbf{x} + \mathbf{\epsilon}$, with $\mathbf{H}\in\mathbb{R}^{m\times n}$ denoting the sensing matrix and $\mathbf{\epsilon}\sim \mathcal{N}(\mathbf{0},\mathbf{I}\sigma^2)$ representing noise. Due to the ill-posed nature of these problems, where solutions may be non-unique or unstable \cite{shanno1978matrix}, recovery is formulated as $\hat{\mathbf{x}} = \operatorname{arg\ min}_{\mathbf{x}} \Big\{ g(\mathbf{x}) + \rho h(\mathbf{x}) \Big\}$,
with a data fidelity term $g(\mathbf{x})=\frac{1}{2}\|\mathbf{H}\mathbf{x}-\mathbf{y}\|_2^2$, and a regularizer $h(\mathbf{x})$ that imposes prior knowledge on the signal such as sparsity \cite{schmidt2005least}, Tikhonov \cite{golub1999tikhonov}, total variation \cite{strong2003edge}, among others.
Beyond hand-crafted regularization, learning based priors have been included into traditional solvers such as FISTA \cite{beck2009fast} or ADMM \cite{admm}. Particularly, Plug-and-Play (PnP) \cite{pnp2013Venkatakrishnan} and regularization by denoising (RED) \cite{RED} have long been used by including a learned-based denoiser as an implicit prior. Recently, deep learning approaches have demonstrated the potential to learn the mapping directly from measurements to reconstructions, bypassing the need for hand-crafted inversion formulas \cite{gualdron2024deep}.

Despite these advances, the ill-posedness of $\mathbf{H}$ often leads to slow convergence and suboptimal recovery. Preconditioning techniques aim to mitigate these issues by reshaping the optimization landscape \cite{nielsen2010efficient}. In traditional linear preconditioning, a fixed operator $\mathbf{P}\in\mathbb{R}^{n\times n}$ is applied in the fidelity gradient update:
\vspace{-0.3cm}
\begin{equation}
\vspace{-0.1cm}
\mathbf{\hat{x}}^k = \mathbf{\hat{x}}^{k-1} - \eta\, \mathbf{P}\Bigl(\overbrace{\mathbf{H}^\top\bigl(\mathbf{H}\mathbf{\hat{x}}^{k-1} - \mathbf{y}\bigr)}^{\nabla g(\mathbf{\hat{x}}^{k-1})}\Bigr),
\label{eq:grad_linear}
\end{equation}
where $\eta$ is the gradient step. Usually, designs for $\mathbf{P}$ are based on the fidelity term Hessian matrix inverse (Newton-like methods) \cite{dassios2015preconditioner,fessler1999conjugate} or polynomial functions of the gram matrix eigenvalues \cite{iyer2024polynomial}. While effective, these linear methods are inherently limited by their dependence on the structure of $\mathbf{H}$ \cite{gander2017origins}.  In contrast, learned preconditioning methods leverage a dataset to design the preconditioning matrix, allowing it to adapt to the specific characteristics of both the data and the inverse problem \cite{ehrhardt2024learning}. While these methods aim to accelerate convergence, minimizing the data fidelity term can introduce a non-trivial null space, which may affect the quality of the final solution and the ability to recover the original signal accurately. 

In this work, we propose D$^2$GP, a framework based on Knowledge Distillation (KD) \cite{hinton2015distilling} ideas to learn a nonlinear preconditioning operator (NPO). Unlike conventional KD, where the teacher model has substantially more trainable parameters than the student, our teacher algorithm is characterized by a better-conditioned sensing matrix and guides a student algorithm that operates under practical, ill-conditioned settings. By leveraging a gradient loss to align data fidelity gradients and an imitation loss to match reconstruction outputs, our approach effectively combines the robustness of classical linear preconditioning with the adaptability of nonlinear methods, yielding enhanced convergence and improved reconstruction performance. This approach benefits from interpretability in the optimization of the preconditioning operator due to distillation losses based on well-posed solutions of the teacher.

\noindent Our main contributions are:
\begin{itemize}
    \vspace{-0.05cm}
    \item We introduce a KD-based method for learning a nonlinear preconditioner for inverse problems.
    \item We propose loss functions that align the gradient directions and outputs between a well-conditioned teacher algorithm and an ill-conditioned, physically-feasible student algorithm.
    \item We validate the proposed framework across multiple imaging modalities, demonstrating performance gains over state-of-the-art preconditioning approaches.
\end{itemize}


\section{D$^2$GP: Deep Distillation Gradient Preconditioning}
We propose D$^2$GP to learn an NPO via a KD framework \cite{suarez2025distilling}. Unlike traditional KD, where the teacher has more parameters, here the teacher algorithm has a better-posed sensing matrix, guiding the student's NPO learning via gradient matching.
\vspace{-0.1cm}
\subsection{Student and Teacher Algorithm Settings} \vspace{-1mm}
Here, the \profe{student algorithm uses} a sensing matrix $\mathbf{H}_s\in\mathbb{R}^{m_s\times n}$ to form measurements $\mathbf{y}_s = \mathbf{H}_s \mathbf{x} + \mathbf{e}_s$.  
The teacher \profe{algorithm uses} impractical, virtual sensing matrix $\mathbf{H}_t\in\mathbb{R}^{m_t\times n}$ ($m_t\gg m_s$) yielding $\mathbf{y}_t = \mathbf{H}_t \mathbf{x} + \mathbf{e}_t$.  
Because $\mathbf{H}_s$ is more ill-posed than $\mathbf{H}_t$, the student relies on NPO to match the teacher's performance.  
We validate this setup on \profe{three tasks}. \vspace{0.1cm}


\noindent \hspace{0.4cm}  $\textbf{1. Single-Pixel Camera (SPC)}$: Both $\mathbf{H}_s\in\{-1,1\}^{m_s\times n}$ and $\mathbf{H}_t\in\{-1,1\}^{m_t\times n}$ select rows from the Hadamard basis \cite{zhang2017hadamard}, but $m_s \ll m_t$ so the student \profe{has} fewer snapshots \profe{available}. Increasing $m$ improves reconstruction at the cost of time. Compression ratios are $\gamma_s = m_s / n$ and $\gamma_t = m_t / n$ \cite{duarte2008single}.
\vspace{0.1cm}

\noindent  \hspace{0.4cm} $\textbf{2. Magnetic Resonance Imaging (MRI)}$:
For single-coil MRI,  
$\mathbf{F} \in \mathbb{C}^{n\times n}$ is the 2D DFT,  
$\mathbf{M}_s \in \{0,1\}^{m_s\times n}$ and $\mathbf{M}_t \in \{0,1\}^{m_t\times n}$ are undersampling masks, so  
$
  \mathbf{H}_s = \mathbf{M}_s\,\mathbf{F} \;\in\;\mathbb{C}^{m_s\times n}, \quad
  \mathbf{H}_t = \mathbf{M}_t\,\mathbf{F} \;\in\;\mathbb{C}^{m_t\times n}.
$
Acceleration factors $AF_s = n / \|\mathbf{M}_s\|_0$ and $AF_t = n / \|\mathbf{M}_t\|_0$ satisfy $AF_s \gg AF_t$. Thus, the student is fast but less accurate, while the teacher is slow yet more precise \cite{pruessmann1998coil}.
\vspace{0.1cm}

\noindent \hspace{0.4cm} $\textbf{3. Super Resolution (SR)}$: Let $\mathbf{D}_s\in\{0,1\}^{m_s\times n}$ and $\mathbf{D}_t\in\{0,1\}^{m_t\times n}$ be downsampling matrices, and $\mathbf{B}_s,\mathbf{B}_t\in\mathbb{R}^{n\times n}$ the blur matrices. Then  
$\mathbf{H}_s=\mathbf{D}_s\mathbf{B}_s\in\mathbb{R}^{m_s\times n}$ and  
$\mathbf{H}_t=\mathbf{D}_t\mathbf{B}_t\in\mathbb{R}^{m_t\times n}$.  
Since $RF_s=\sqrt{n/m_s}\gg RF_t=\sqrt{n/m_t}$, the student works with feasible low-res inputs; the teacher requires high-res for better performance \cite{yang2010image}. 

\vspace{-0.1cm}
\subsection{Nonlinear preconditioning operator structure}

The NPO $\mathcal{P}_\theta: \mathbb{R}^{n} \rightarrow \mathbb{R}^{n} $ has trainable parameters $\theta$ and is implemented as a neural network. In Section \ref{sec:results} an ablation study is presented with several neural network architectures. Additionally, we consider a strategy so that the NPO adapts to every algorithm iteration. We employ a positional encoding method based on the algorithm iteration. Mainly, $\mathcal{P}_\theta(\nabla g_s(\hat{\mathbf{x}}^k),\phi(k))$ where $\phi(k) = \mathbf{w} k $ where $\mathbf{w} \in \mathbb{R}^d$  is an embedding function with $d$ number of features. This embedding vector is also optimized. For simplicity, we consider that $\theta$ has both the embedding parameter and the model weights. 
\vspace{-0.4cm}
\subsection{Distilled preconditioning operator}
Our approach is data-driven, where a dataset of $B$ clean images $\mathcal{X} = \{\mathbf{x}_i\}_{i=1}^{B}$ is employed to generate the measurement sets as $\mathcal{Y}_s = \{\mathbf{y}_{s_i} \vert \mathbf{y}_{s_i} = \mathbf{H}_s\mathbf{x}_i + \mathbf{e}_{s_i}\}_{i=1}^B$ and  $\mathcal{Y}_t = \{\mathbf{y}_{t_i} \vert \mathbf{y}_{t_i} = \mathbf{H}_t\mathbf{x}_i + \mathbf{e}_{t_i}\}_{i=1}^B$. The optimization of the NPO is proposed as 
\vspace{-0.2cm}
\begin{align}
\vspace{-0.3cm}
   {{\mathbf{\theta}}^{\star}}=\ &\underset{{{\mathbf{\theta}}}}{\operatorname{arg\ min}} \ 
   \mathbb{E}_{\mathbf{x}_i,\mathbf{y}_{s_i},\mathbf{y}_{t_i}}
    \mathcal{L}_{KD}\left(\hat{\mathbf{x}}_{\mathcal{P}_i},{\mathbf{H}_s}, {\mathbf{H}_t},{\hat{\mathbf{x}}_{t_i}}, {\mathcal{P}_{\mathbf{\theta}}}\right),\nonumber\\
    &\text{ s.t }\hat{\mathbf{x}}_{\mathcal{P}_i}=\texttt{\profe{Algorithm 1}}\left({\mathcal{P}_{\mathbf{\theta}}},{\mathbf{H}_s},{\mathbf{y}_{s_i}}\right)\nonumber\\
    &\hphantom{\text{ s.t }}\hat{\mathbf{x}}_{t_i}=\texttt{\profe{Algorithm 1}}\left(\mathcal{I},{\mathbf{H}_t},{\mathbf{y}_{t_i}}\right),\label{eq:opt_P_theta}
\end{align}
\noindent \profe{where Algorithm \ref{alg:npo_fista} refers to the Preconditioned PnP-FISTA, which is used for both, the gradient-preconditioned student algorithm $\texttt{GPSA} := \texttt{Alg1}\left({\mathcal{P}_{\mathbf{\theta}}},{\mathbf{H}_s},{\mathbf{y}_{s_i}}\right)$ and the teacher algorithm $\texttt{TA} := \left(\mathcal{I},{\mathbf{H}_t},{\mathbf{y}_{t_i}}\right)$, with $\mathcal{I}$ as an identity function. Eq. \eqref{eq:opt_P_theta} is a bilevel optimization problem \cite{colson2007overview}, where the $\texttt{TA}$ is used to distill its knowledge into the NPO weights, $\theta$, thereby improving the $\texttt{GPSA}$.} Here,  the cost function $\mathcal{L}_{KD}(\cdot)$ is crucial, and depends on the teacher and student sensing matrices and their reconstructions\profe{, defined as}
\vspace{-0.2cm}
\begin{align}
\vspace{-0.2cm}
    &\mathcal{L}_{KD}(\hat{\mathbf{x}}_{\mathcal{P}_i},{\mathbf{H}_s}, {\mathbf{H}_t},{\hat{\mathbf{x}}_{t_i}}, {\mathcal{P}_{\mathbf{\theta}}})= \lambda_G \mathcal{L}_{G} + \lambda_I \mathcal{L}_I + \lambda_S \mathcal{L}_S. 
\end{align}

\textbf{Gradient loss} ($\mathcal{L}_G$): Aims to align the direction of the $\texttt{GPSA}$'s {with the} $\texttt{TA}$'s data fidelity gradients as follows
\vspace{-0.2cm}
\begin{equation}
\vspace{-0.2cm}
\mathcal{L}_{G}=\Bigg\vert 1- \mathcal{S}_c\bigg({\mathcal{P}_{\mathbf{\theta}}}\Big( \nabla g_s(\hat{\mathbf{x}}_{s_i}), \phi(K)\Big), \nabla g_t({\hat{\mathbf{x}}_{t_i}})\bigg)\Bigg\vert^2, \label{eq:LG}
\end{equation}
where $g_s(\hat{\mathbf{x}}_{s_i}) = \Vert\mathbf{y}_{s_i}-\mathbf{H}_s\hat{\mathbf{x}}_{s_i}\Vert_2^2$ and  $g_t(\hat{\mathbf{x}}_{t_i}) = \Vert\mathbf{y}_{t_i}-\mathbf{H}_t\hat{\mathbf{x}}_{t_i}\Vert_2^2$ are the student and teacher data fidelity terms, $\mathcal{S}_c$ is the cosine similarity, i.e., $\mathcal{S}_c(\mathbf{a},\mathbf{b}) = \frac{{\mathbf{a}^\top\mathbf{b}}}{\Vert\mathbf{a}\Vert\Vert\mathbf{b}\Vert}$. 

\textbf{Imitation loss} ($\mathcal{L}_I$): Its main objective is to ensure that the $\texttt{GPSA}$ produces the same output as the $\texttt{TA}$, defined as 
\vspace{-0.2cm}
\begin{equation}
\vspace{-0.2cm}
   \mathcal{L}_I = \Vert\hat{\mathbf{x}}_{\mathcal{P}_i}-{\hat{\mathbf{x}}_{t_i}}\Vert_2^2.\label{eq:LI}
\end{equation}

\textbf{Supervised loss} ($\mathcal{L}_{S}$):
Optimizes $\texttt{GPSA}$'s NPO using the ground truth label $\mathbf{x}_i$ instead of the teacher's output, as 
\vspace{-0.2cm}
\begin{equation}
\vspace{-0.2cm}
\mathcal{L}_{S}= \Vert\hat{\mathbf{x}}_{\mathcal{P}_i}-{{\mathbf{x}_i}}\Vert_2^2.    
\end{equation}

\noindent Combining these loss functions allows the NPO design to behave like the $\texttt{TA}$ in the $\texttt{GPSA}$ without being highly affected by the physical limitations of the real implementation. As detailed in Algorithm~\ref{alg:kd_npo}, the optimization problem is solved using off-the-shelf stochastic gradient descent (GD) algorithms such as Adam \cite{Adam} or AdamW \cite{loshchilov2017decoupled}. Note that, unlike traditional preconditioning operator design, which only aims to improve the algorithm's convergence rate,  our approach aims to improve recovery performance through $\texttt{TA}$ guidance using the $\mathcal{L}_I$ loss function and to achieve a fast convergence rate via $\mathcal{L}_G$.

\begin{algorithm}[!t]
\footnotesize 
\caption{Preconditioned PnP-FISTA}
\label{alg:npo_fista}
\begin{algorithmic}[1]
\Require Preconditioner $\mathcal{P}$, Sensing matrix $\mathbf{H}$, Measurements $\mathbf{y}$, 
\textcolor{white}{Require: Requ} Step size $\alpha$, Regularization weight $\rho$, Iterations $K$.
  \State $\mathbf{x}^0 = \mathbf{z}^1 \gets \mathbf{0},\quad t^{0} \gets 1$
  \Comment{Initializations}
  \For{$k=1,\dots,K$}
    \State $\mathbf{x}^k \gets \mathbf{z}^k - \alpha\,\mathcal{P}\bigl(\nabla g(\mathbf{z}^k),\textcolor{black}{\phi(k)}\bigr)$ 
    \Comment{Preconditioned GD }
    \State $\mathbf{x}^k \gets \operatorname{prox}_{\alpha\rho h}\bigl(\mathbf{x}^k\bigr)$ 
    \Comment{Proximal mapping}
    \State $t^{k} \gets \tfrac{1 + \sqrt{1 + 4(t^{k-1})^2}}{2}$ 
    \Comment{Update Nesterov momentum}
    \State $\mathbf{z}^{k+1} \gets \mathbf{x}^k + \tfrac{t^{k-1} - 1}{t^{k}}\bigl(\mathbf{x}^k - \mathbf{x}^{k-1}\bigr)$ 
  \EndFor
  \State \Return $\mathbf{x}^K$ 
  \Comment{Final reconstruction}
\end{algorithmic}
\end{algorithm}

\begin{algorithm}[!t]
\footnotesize 
\caption{KD-based Training of Nonlinear Preconditioner}
\label{alg:kd_npo}
\begin{algorithmic}[1]
\Require Images $\mathcal{X}$, Measurements $(\mathcal{Y}_s,\mathcal{Y}_t)$, Sensing matrices $(\mathbf{H}_s, \mathbf{H}_t)$, \textcolor{white}{Requ} Step sizes $(\alpha_s, \alpha_t)$, Iterations $K$, Epochs $N$, Learning rate $\eta$, \textcolor{white}{Requie} \hspace{-0.25cm} Coefficients $(\lambda_G, \lambda_I, \lambda_S)$, Regularization weight $\rho$, Batches $B$.
\State $\theta \sim \mathcal{N}(0,\sigma^2)$ \Comment{Initialize NPO's weights}
\For{$epoch = 1,\dots,N$}
  \For{ $i = 1,\dots, B$}
  \State $\mathbf{x}_i \gets \mathcal{X}[i],\quad \mathbf{y}_{s_i}\gets \mathcal{Y}_s[i],\quad \mathbf{y}_{t_i}\gets \mathcal{Y}_t[i]$ 
  \State $\hat{\mathbf{x}}_{t_i} = \texttt{Alg1} (\mathcal{I}, \mathbf{H}_t,\mathbf{y}_{t_i}, \alpha_t,\rho,K)$ \Comment{$\texttt{TA}$ recovery}
    \State $\hat{\mathbf{x}}_{\mathcal{P}_i} = \texttt{Alg1} (\mathcal{P}_\theta, \mathbf{H}_s,\mathbf{y}_{s_i}, \alpha_s,\rho,K)$  \Comment{$\texttt{GPSA}$ recovery}
    
    \State $\mathcal{L}_G=\left|1-\tfrac{\mathcal{P}_\theta(\nabla g_s(\hat{\mathbf{x}}_{\mathcal{P}_i} ), \phi({K}))^\top \nabla {g}_t(\hat{\mathbf{x}}_{t_i} )}{\|\mathcal{P}_\theta(\nabla g_s(\hat{\mathbf{x}}_{\mathcal{P}_i} ), \phi({K}))\|\|\nabla {g}_t(\hat{\mathbf{x}}_{t_i} )\|}\right|^2$ \Comment{Gradient loss}
    \State $\mathcal{L}_I=\|\hat{\mathbf{x}}_{\mathcal{P}_i} -\hat{\mathbf{x}}_{t_i}\|_2^2$ \Comment{Imitation loss} \vspace{1mm}
    \State $\mathcal{L}_S=\|\hat{\mathbf{x}}_{\mathcal{P}_i} -\mathbf{x}_i\|_2^2$ \Comment{Supervised loss} \vspace{1mm}
    \State $\theta\gets\theta-\eta\,\nabla_\theta\bigl(\lambda_G\mathcal{L}_G+\lambda_I\,\mathcal{L}_I+\lambda_S\,\mathcal{L}_S\bigr)$
  \EndFor
\EndFor
\State\Return $\mathcal{P}_{\mathbf{\theta}}$ \Comment{KD-optimized NPO}
\end{algorithmic}

\end{algorithm}

\section{Experimental Setup} \label{sec:results}
The proposed approach was validated across three common imaging modalities: SPC, MRI, and SR; using the PnP implementation and \textcolor{black}{the DnCNN pretrained denoiser} from the DeepInv Library \cite{Tachella_DeepInverse_A_deep_2023} for image recovery. {{We used the FISTA algorithm \cite{beck2009fast}; however, the proposed approach can be extended to other algorithms.} The stepsizes are $\alpha_t = 0.7$ and $\alpha_s = 0.4$ with $K=20$ iterations. The NPO was optimized for $N=50$ epochs with a learning rate $\eta=1\times 10^{-5}$. The computer used has an Intel(R) Xeon(R) W-3223 CPU @ 3.50GHz, 100 GB RAM, and an NVIDIA Quadro RTX 8000 with 48 GB VRAM.

\textbf{SPC:} {We use the MNIST dataset \cite{deng2012mnist},} with $50k$ training and $10k$ testing images, all resized to $32 \times 32$, $n=1024$. Adam \cite{Adam} optimizer was used with a $225$ batch size. A 2D subsampled Hadamard basis was used for the sensing matrix. The compression ratios are $\gamma_s=0.2$ and $\gamma_t=0.7$.

\textbf{MRI:} We use the FastMRI single-coil knee dataset from \cite{FastMRI_dataset}, preprocessed by \cite{Tachella_DeepInverse_A_deep_2023}. This dataset consists of $900$  training and  $73$  testing MRI knee images, all resized to $50 \times 50$, $n=2500$. The AdamW optimizer \cite{loshchilov2017decoupled}  was employed with a weight decay of $0.01$. The batch size was set to $32$. For all MRI experiments, we employed a 1D Gaussian undersampling mask. The acceleration factors are $AF_s=5$ and $AF_t=1$.

\textbf{SR:} We used the CelebA dataset \cite{liu2015faceattributes}, with $162k$ training and $19k$ testing images, all resized to $110 \times 110$, $n=12100$. Adam optimizer \cite{Adam} was used. The batch size was $3$. Here, $\mathbf{B}_s=\mathbf{B}_t$ is the convolution matrix of a $9 \times 9$ Gaussian filter with $\sigma=1$. The resolution factors are $RF_s=4$ and $RF_t=1$.

\begin{figure}[!b]
  \centering
    \includegraphics[width=\linewidth]{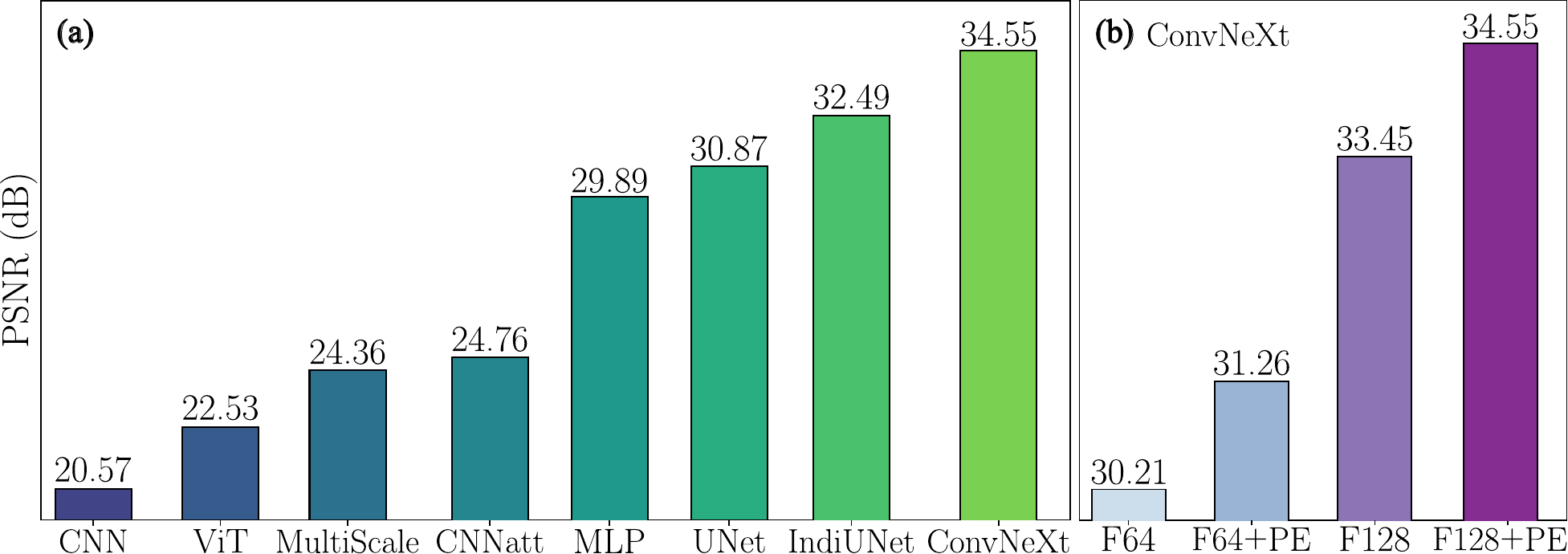}
    \vspace{-0.7cm}
    \caption{\hspace{2mm} (a) Ablation results in terms of PSNR of different NNs as NPO for SPC. (b) Number of features and positional encoding usage in ConvNeXt.}
    \label{fig:networks_ablation}
    \vspace{-0.3cm}
\end{figure}

\subsection{Neural Network ablation studies}
{This experiment aims to identify the best neural network architecture for the NPO.} {We evaluated the following neural networks (NNs) architectures: } Multilayer perceptron \cite{rosenblatt1958perceptron}, convolutional NN (CNN) \cite{lecun1998gradient}, CNN with attention module \cite{woo2018cbam}, UNet \cite{ronneberger2015u}, MultiScale CNN \cite{yuan2018multiscale}, IndiUNet \cite{delbracio2023inversion}, Vision Transformer (ViT) \cite{dosovitskiy2020image}, and ConvNext \cite{liu2022convnet}. For the ablation studies on Fig. \ref{fig:networks_ablation}(a), modifications were made to reduce the number of parameters of each of the NNs, to have fewer parameters than $\mathbf{P}\in\mathbb{R}^{n \times n}$, and selecting the configuration with best possible performance. Due to the parameter reduction, some networks, such as ViT, despite their widely known learning capabilities, are not able to generalize the gradient mapping from student to teacher. Regardless of the configuration, the MLP has more parameters than $\mathbf{P}$ but does not perform adequately. ConvNeXt demonstrated improved performance with fewer parameters due to its use of CNNs.

\begin{figure}[!b]
    \centering
     \includegraphics[width=\linewidth]{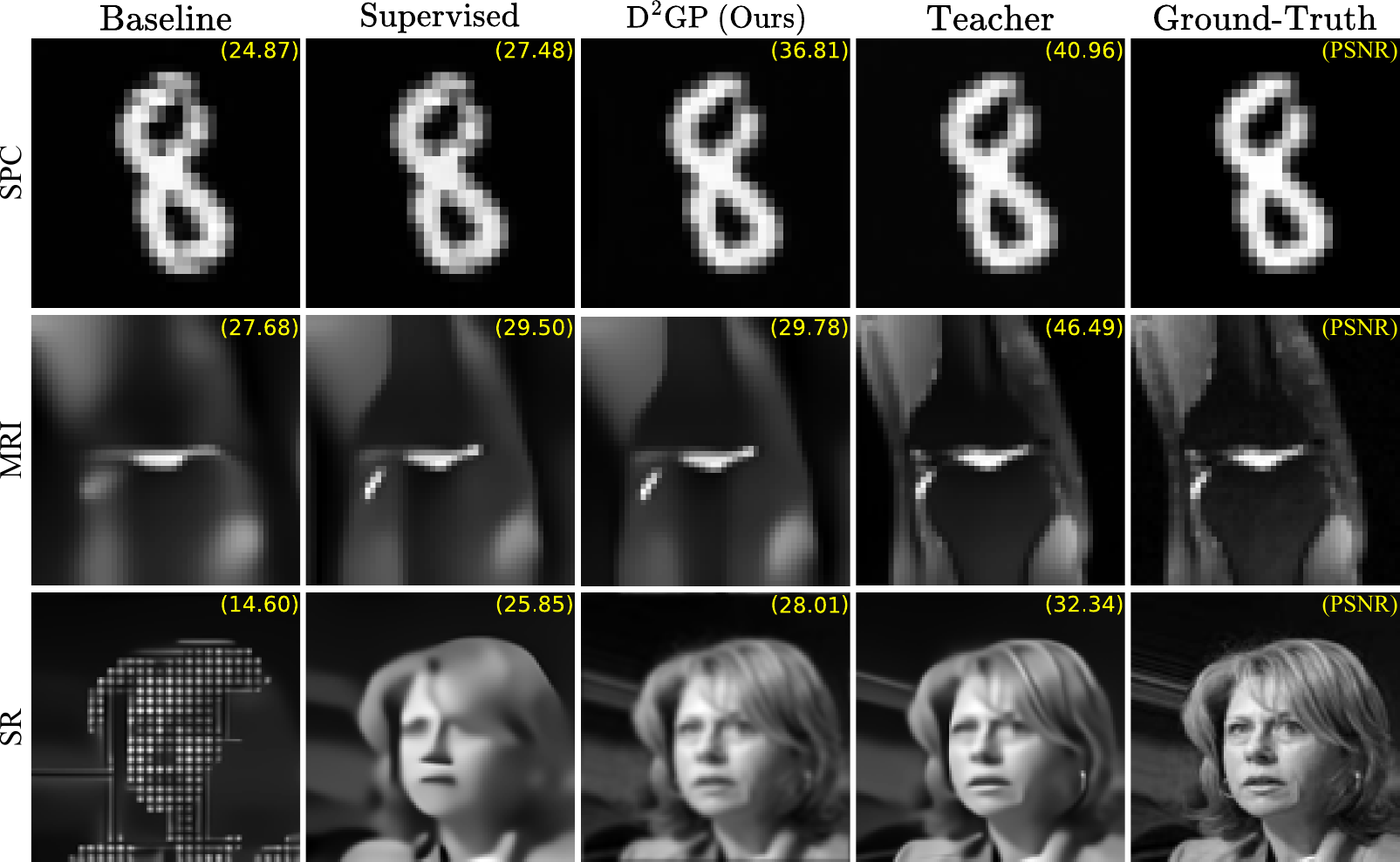}
     \vspace{-0.7cm}
    \caption{Visual results and PSNR for PnP-FISTA using the nonlinear preconditioning methods across SPC, MRI, and SR tasks. 
    } 
    \label{fig:D2GP_VisualResults}
\end{figure}

\subsection{Nonlinear preconditioning operator setup}
We set the number of features of the PE ${d}=256$. The maximum number of features where the efficiency was maintained and the best performance was achieved is 128. Finally, a ConvNeXt network with 5 blocks was used, with a ReduceLROnPlateau scheduler \cite{al2022scheduling}, a residual connection in the last layer, and the PE proposed in IndiUNet \cite{delbracio2023inversion} so that the network could know and differentiate the gradients according to the current iteration of the algorithm. 

\begin{table*}[!t]
\centering
\caption{SOTA comparison across Single-Pixel Camera (SPC), Magnetic Resonance Imaging (MRI), and Super-Resolution (SR) tasks.}
\vspace{-0.3cm}
\resizebox{\linewidth}{!}{
\begin{tabular}{c|c|c|ccc|ccc|ccc}
\hline\hline
\multirow{2}{*}{\textbf{Learned}} 
  & \multirow{2}{*}{\textbf{Method}} 
    & \multirow{2}{*}{\textbf{Formulation}} 
      & \multicolumn{3}{c|}{\textbf{SPC} ($\gamma_s=0.2$, $\gamma_t=0.7$)} 
        & \multicolumn{3}{c|}{\textbf{MRI} ($AF_s=5$, $AF_t=1$)} 
          & \multicolumn{3}{c}{\textbf{SR} ($RF_s=4$, $RF_t=1$)} \\
\cline{4-12}
& & 
  & PSNR & Params & Ratio 
    & PSNR & Params & Ratio 
      & PSNR & Params & Ratio \\
\hline
\xmark & Baseline & $\mathcal{I}( \nabla g_s)$ & 22.36 & 0 & 0 & 25.77 & 0 & 0 & 11.14 & 0 & 0 \\
\xmark & Hessian~\cite{fessler1999conjugate} & $(\mathbf{H}_s^\top\mathbf{H}_s)^{-1}\nabla g_s$ & 22.94 & 0 & 0 & 28.16 & 0 & 0 & 11.75 & 0 & 0 \\
\xmark & Polynomial~\cite{iyer2024polynomial} & $p(\mathbf{H}_s^\top\mathbf{H}_s) \nabla g_s$ & 22.01 & 5 & $6 \times 10^{-6}$ & 28.07 & 5 & $3 \times 10^{-6}$ & 11.81 & 5 & $6 \times 10^{-6}$ \\
\cmark & Scalar step~\cite{ehrhardt2024learning} & $p_k \nabla g_s$ & 23.36 & 20 & $2 \times 10^{-5}$ & 27.88 & 20 & $1 \times 10^{-5}$ & 11.62 & 20 & $2 \times 10^{-5}$ \\
\cmark & Convolutional~\cite{ehrhardt2024learning} & $\mathbf{p_k}  \ast \nabla g_s$ & 21.30 & 500 & $6 \times 10^{-4}$ & 28.63 & 500 & $3 \times 10^{-4}$ & 17.55 & 500 & $6 \times 10^{-4}$ \\
\cmark & Pointwise~\cite{ehrhardt2024learning} & $\mathbf{p_k \odot} \nabla g_s$ & 21.28 & $20k$ & $0.024$ & 27.59 & $50k$ & $0.03$ & 11.19 & $242k$ & $0.28$ \\
\cmark & Full-linear~\cite{ehrhardt2024learning} & $\mathbf{P_k} \nabla g_s$ & 26.19 & $21M$ & $24.42$ & 28.15 & $125M$ & $72.8$ & 22.02 & $2928M$ & $3410$ \\
\cmark & Supervised & $\mathcal{P}_{\bm{\theta}_{Sup}^{\star}}\left(\nabla g_s, \phi(k)\right)$ & \underline{33.79} & $858k$ & $1$ & \underline{28.87} & $1.7M$ & $1$ & \underline{25.26} & $858k$ & $1$ \\
\cmark & D$^2$GP (Ours) & $\mathcal{P}_{\bm{\theta}_{KD}^\star}\left(\nabla g_s, \phi(k)\right)$ & \textbf{34.55} & $858k$ & $1$ & \textbf{29.39} & $1.7M$ & $1$ & \textbf{25.88} & $858k$ & $1$ \\
 \hline \hline
\end{tabular}
}
\label{tab:SOTA_D2GP_AllTasks}
\vspace{-0.4cm}
\end{table*}

\begin{figure*}[!ht]
    \centering
    \includegraphics[width=0.92\linewidth]{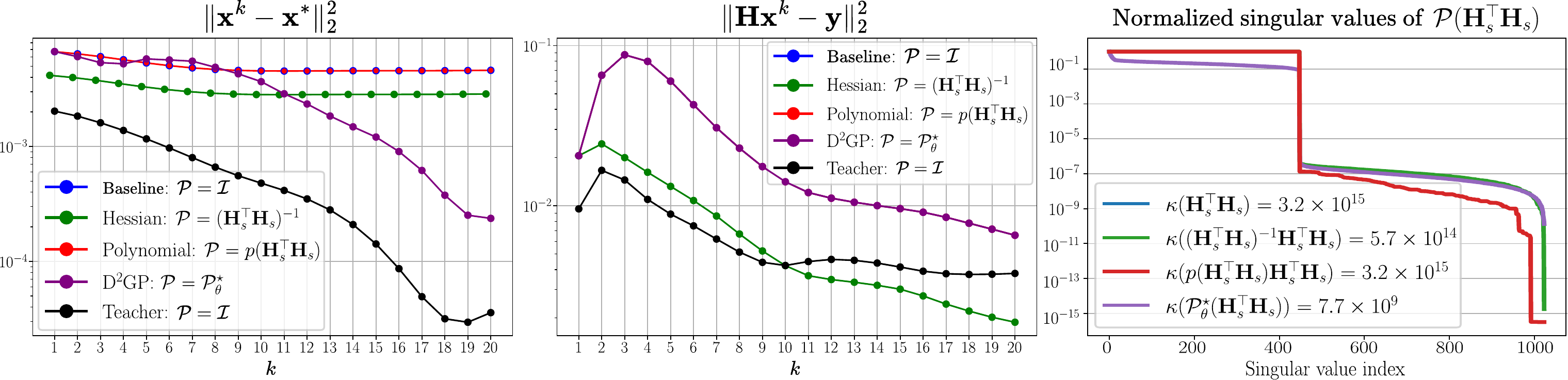}
    \vspace{-0.4cm}
    \caption{(a) Reconstruction convergence along algorithm iterations. (b) Data fidelity term convergence. (c) Preconditioned Gram matrix's singular values and condition number $\kappa$ for SPC with D$^2$GP and the most parameter-efficient SOTA methods. For D$^2$GP, a linear approximation was applied.}
    \label{fig:D2GP_analysis}
    \vspace{-0.4cm}
\end{figure*}



\subsection{State-of-the-art (SOTA) comparison}
An extensive comparison was conducted among state-of-the-art methods, divided between those that require learning through optimization 
\profe{and those that do not.}
In addition to the proposed method D$^2$GP, the formulation that uses only $\mathcal{L}_{S}$ (no teacher) is named Supervised. Table \ref{tab:SOTA_D2GP_AllTasks} summarizes PSNR results, the number of trainable parameters, and the parameter ratio \textcolor{black}{$\texttt{num\_params}\text{(Method)}/\texttt{num\_params}\text{(D}^2\text{GP)}$} for different SOTA methods preconditioning \textcolor{black}{PnP-FISTA}. In SPC, D$^2$GP outperforms the best SOTA method by over 8 dB while using 24 times fewer parameters. The largest parameter reduction occurs in SR, achieving 3 dB gains with 3410 times fewer parameters than Full-linear, highlighting D$^2$GP’s invariance to image dimensionality thanks to its ConvNeXt-based architecture.

Fig. \ref{fig:D2GP_VisualResults} shows visual results for the SPC, MRI, and SR tasks, where the proposed method outperforms the supervised method, bounded by the Teacher performance. Furthermore, although some methods require significantly fewer parameters than D$^2$GP, such as the unlearnable or parameterized ones, their performance is very poor. Thus, despite their efficiency, they significantly compromise reconstruction quality.


\subsection{Local Linearization via the Jacobian}
To analyze the eigenvalues of the preconditioning operator, we sought a linear approximation, which, although not precise, enables us to examine its condition number. \noindent First, let $\mathcal{P}_{\bm{\theta}^\star} : \mathbb{R}^n \to \mathbb{R}^n$ be a differentiable, pretrained nonlinear preconditioner. In a small neighborhood around $\mathbf{x}_0$, its behavior is well-approximated by its Jacobian,
$
\left(J_{\mathcal{P}_{\bm{\theta}^\star}}(\mathbf{x}_0)\right)_{ij} = \frac{\partial (\mathcal{P}_{\bm{\theta}^\star}(\mathbf{x}_0))_i}{\partial x_j}.
$
Thus, for a small perturbation $\boldsymbol{\delta}$, we have
$
\mathcal{P}_{\bm{\theta}^\star}(\mathbf{x}_0+\boldsymbol{\delta}) \approx \mathcal{P}_{\bm{\theta}^\star}(\mathbf{x}_0) + J_{\mathcal{P}_{\bm{\theta}^\star}}(\mathbf{x}_0)\,\boldsymbol{\delta},
$
where $\delta \in \mathbb{R}^n$ denotes an arbitrary small perturbation (displacement) vector around $\mathbf{x}_0$. In effect, we obtain an approximate impulse response for each entry by injecting canonical unit vectors into the network. Since the analytical Jacobian is typically unavailable in deep networks, we estimate it via finite differences. For a small finite‑difference step size $\epsilon>0$, the partial derivative with respect to $x_j$ is approximated as
$
\frac{\partial (\mathcal{P}_{\bm{\theta}^\star}(\mathbf{x}_0))_i}{\partial x_j} \approx \frac{(\mathcal{P}_{\bm{\theta}^\star}(\mathbf{x}_0+\epsilon\,\mathbf{e}_j))_i - (\mathcal{P}_{\bm{\theta}^\star}(\mathbf{x}_0))_i}{\epsilon},
$
where $\mathbf{e}_j$ is the $j$th standard basis vector in $\mathbb{R}^n$. Repeating this process for all components yields an effective approximation of the Jacobian~\cite{raghu2017expressive,dennis1996numerical}.   In Fig. \ref{fig:D2GP_analysis}, we observe the convergence of the reconstruction concerning the GT, the fidelity term, and the singular values of the Gram matrix for SPC. The proposed D$^2$GP method exhibits the fastest convergence over 20 iterations of the PnP-FISTA algorithm, reaching comparable performance against the $\texttt{TA}$ in fewer iterations. Although its fidelity term convergence is not the best, this does not hinder the overall reconstruction quality, as methods with lower fidelity tend to converge to suboptimal local minima. 
 In summary, the proposed method achieves faster convergence and better reconstruction quality. Fig. \ref{fig:NPOs} shows a zoomed version of the linearization of $\mathcal{P}_{\bm{\theta}_{KD}^\star}$, where the closeness to an identity matrix, the importance of the diagonal, and the variability depending on the task and its parameters are observed.

\begin{figure}[!b]
    \centering
    \includegraphics[width=\linewidth]{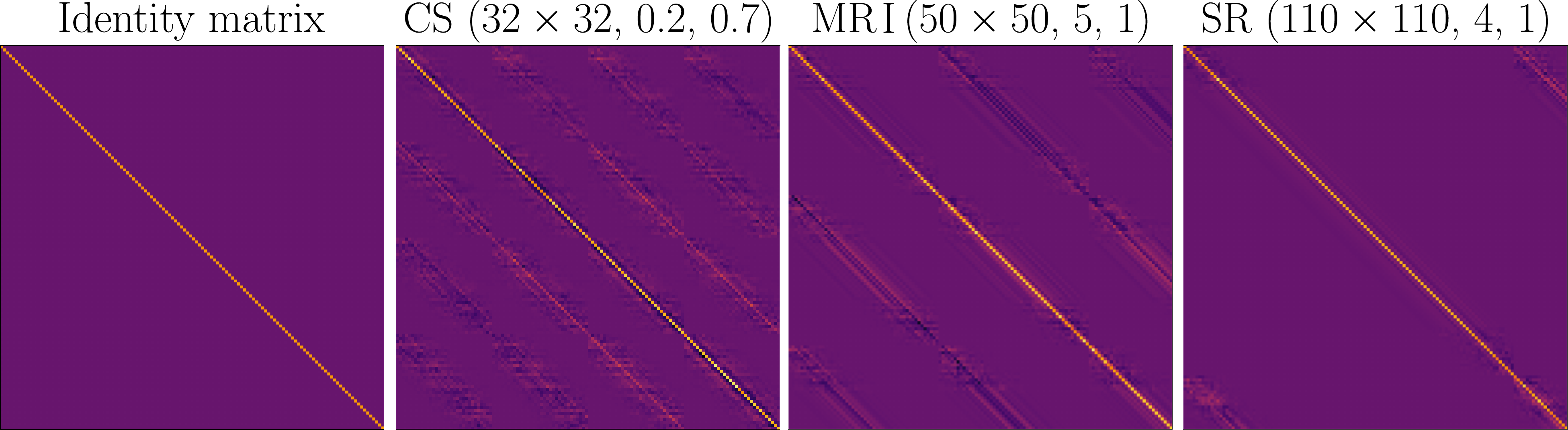}
    \vspace{-0.7cm}
    \caption{\textcolor{black}{$128 \times 128$ zoomed version of the linear approximation of the learned NPO $\mathcal{P}_{\bm{\theta}_{KD}^\star}$.}}
    \label{fig:NPOs}
    \vspace{-0.2cm}
\end{figure}

\section{\textcolor{black}{Discussion  and Future Work}}

The proposed method improves efficiency and performance over linear and supervised preconditioners, yet its conditioning analysis relies on an approximate linearization of the NPO, which may not be exact. Despite this, experimental results confirm the method’s effectiveness across tasks. Future work includes exploring alternative loss functions, distillation strategies, higher-resolution tasks, real-world validation, and a more precise theoretical analysis of the induced conditioning and teacher settings.

\section{Conclusions}
We introduced a KD-based preconditioning method that enhances convergence speed and reconstruction quality using convolutional architectures. Unlike traditional linear and supervised approaches, our method guides the student algorithm with a better-conditioned teacher, leading to superior performance. By combining the generalization and robustness of KD with the efficiency of gradient preconditioning, our approach achieves faster, more accurate, and parameter-efficient optimization. This results in a practical strategy that consistently improves convergence and reconstruction across diverse inverse imaging tasks, without scaling with image dimensions.

\clearpage
\balance
\bibliographystyle{IEEEbib}
\bibliography{refs}

\end{document}